\documentclass[reprint, showpacs, amsmath, amssymb, aps, prb, superscriptaddress]{revtex4-1}

\usepackage{xcolor}
\usepackage{graphicx}
\usepackage{dcolumn}% Align table columns on decimal point
\usepackage{bm}% bold math
\usepackage{hyperref}% add hypertext capabilities
\usepackage{siunitx}
\usepackage{todonotes}
\usepackage{cleveref}
\usepackage{amsmath,amssymb}
\usepackage[normalem]{ulem}

\graphicspath{{figures/}}

\begin{document}

\title{Comparison of temperature and doping dependence of nematic susceptibility near a putative nematic quantum critical point}

\author{J. C. Palmstrom}
\affiliation{Geballe Laboratory for Advanced Materials and Department of Applied Physics, Stanford University, California 94305, USA}
\affiliation{Stanford Institute for Materials and Energy Sciences, SLAC National Accelerator Laboratory, 2575 Sand Hill Road, Menlo Park, CA 94025, USA}

\author{P. Walmsley}
\affiliation{Geballe Laboratory for Advanced Materials and Department of Applied Physics, Stanford University, California 94305, USA}
\affiliation{Stanford Institute for Materials and Energy Sciences, SLAC National Accelerator Laboratory, 2575 Sand Hill Road, Menlo Park, CA 94025, USA}

\author{J. A. W. Straquadine}
\affiliation{Geballe Laboratory for Advanced Materials and Department of Applied Physics, Stanford University, California 94305, USA}
\affiliation{Stanford Institute for Materials and Energy Sciences, SLAC National Accelerator Laboratory, 2575 Sand Hill Road, Menlo Park, CA 94025, USA}

%\author{S. T. Hannahs}
%\affiliation{National High Magnetic Field Laboratory, DC Field Facility, Tallahassee FL, USA}

\author{M. E. Sorensen}
\affiliation{Stanford Institute for Materials and Energy Sciences, SLAC National Accelerator Laboratory, 2575 Sand Hill Road, Menlo Park, CA 94025, USA}
\affiliation{Geballe Laboratory for Advanced Materials and Department of Physics, Stanford University, California 94305, USA}

\author{S. T. Hannahs}
\affiliation{National High Magnetic Laboratory, Florida State University, Tallahassee, FL 32306}

\author{D. H. Burns}
\affiliation{Department of Geological Sciences, Stanford University, Stanford, California 94305, USA}

\author{I. R. Fisher}
\affiliation{Geballe Laboratory for Advanced Materials and Department of Applied Physics, Stanford University, California 94305, USA}
\affiliation{Stanford Institute for Materials and Energy Sciences, SLAC National Accelerator Laboratory, 2575 Sand Hill Road, Menlo Park, CA 94025, USA}

%\author{authors}
%\affiliation{Geballe Laboratory for Advanced Materials and Department of Applied Physics, Stanford University, California 94305, USA}
%\affiliation{Stanford Institute for Materials and Energy Sciences, SLAC National Accelerator Laboratory, 2575 Sand Hill Road, Menlo Park, CA 94025, USA}

\date{\today}

\maketitle
\textbf{Strong electronic nematic fluctuations have been discovered near optimal doping for several families of Fe-based superconductors\cite{Kuo2016}, motivating the search for a possible link between these fluctuations, nematic quantum criticality, and high temperature superconductivity. Here we probe a key prediction of quantum criticality, namely power law dependence of the associated nematic susceptibility as a function of composition and temperature approaching the compositionally-tuned putative quantum critical point. To probe the `bare' quantum critical point requires suppression of the superconducting state, which we achieve by using  large magnetic fields, up to 45 T, while performing elastoresistivity measurements to follow the nematic susceptibility. We performed these measurements for the prototypical electron-doped pnictide, Ba(Fe$_{1-x}$Co$_x$)$_2$As$_2$, over a dense comb of dopings. We find that close to the putative quantum critical point, the nematic susceptibility appears to obey power law behavior over almost a decade of variation in composition, consistent with basic notions of nematic quantum criticality. Paradoxically, however, we also find that the temperature dependence for compositions close to the critical value cannot be described by a single power law. This is surprising as power law scaling in both doping and temperature is expected close to a quantum critical point \cite{Sachdev2011a}.
 }

%Quantum critically empirically seems to be closely linked to unconventional superconductivity in many materials including heavy Fermion superconductors \textcolor{red}{ref} and, albeit less clearly, cuprate and Fe-based materials \textcolor{red}{ref}. There is strong evidence for criticality in the isovalently substituted Ba(Fe$_{1-x}$P$_x$)$_2$As$_2$ \cite{Jiang2009, Nakai2010, Shishido2009, Iye2012, Walmsley, Analytis2014, Shibauchi2014a, Hayes2016}, though the universality of quantum criticality has not been established. It remains an open question for most Fe-based superconductors whether there is avoided criticality\cite{Luo2012, Lu2013}, or one or two quantum critical points 'hidden' beneath the superconducting dome. Ba(Fe$_{1-x}$Co$_x$)$_2$As$_2$ is a representative electron-doped system that has been extensively studied\cite{Ning2010, Yoshizawa2012a, Chu2012, Gallais2013, Bohmer2014a, Kuo2016, Kretzschmar2016}, but disorder due to doping precludes the quantum oscillation measurements which were used to establish quantum criticality in the isovalent system \textcolor{red}{ref}. 

A connection between superconductivity and \emph{magnetic} quantum criticality has been established for a number of heavy fermion systems \cite{Gegenwart2008}. Tentative signatures of the effects of possible quantum phase transitions, such as renormalization of the quasiparticle effective mass, have been found for some cuprate superconductors\cite{Ramshaw2015, Michon2019}, but the situation is less clear due to the possible presence and interaction of multiple nearby electronic phases. Compared to cuprates, the situation in the Fe-based materials is much clearer since the symmetry of the ordered phases is well understood and the phase transitions are clearly identified. There is strong evidence for mass renormalization approaching a possible quantum critical point in isovalently substituted Ba(Fe$_{1-x}$P$_x$)$_2$As$_2$ \cite{ Shishido2009, Hashimoto2012a, Walmsley, Analytis2014}, but to date power law scaling of neither the magnetic nor nematic susceptibility has been observed as a function of composition, and despite suggestive signatures, the universality of quantum criticality has not been established. Indeed, it remains an open question for most Fe-based superconductors whether there is avoided criticality\cite{Luo2012, Lu2013, Reiss2019}, or one or two quantum critical points `hidden' beneath the superconducting dome. The two candidate quantum critical points are a nematic quantum critical point which would have associated rotational symmetry breaking fluctuations and an antiferromagnetic critical point with associated spin-fluctuations. Here, we specifically focus on nematic fluctuations and the variation of the nematic susceptibility upon approach to the associated putative quantum critical point since this is the first of the two possible quantum critical points that are encountered upon approaching the ordered states from the overdoped (tetragonal and non-magnetic) regime. Ba(Fe$_{1-x}$Co$_x$)$_2$As$_2$ was chosen as a representative electron-doped system since the crystal growth of this material is very well controlled, and it is possible to prepare closely spaced compositions spanning the compositionally-tuned phase diagram --- a key requirement for any test of power law behavior. 

  From a theoretical perspective nematic fluctuations have been shown to enhance \emph{any} symmetry of existing superconducting pairing interactions\cite{Maier2014, Metlitski2015, Lederer2015} and even induce superconductivity\cite{Lederer2017}. The observation of strong nematic fluctuations in the high temperature phase of optimally doped Fe-based superconductors\cite{Kuo2016} is consistent with the presence of a nematic quantum critical point, but alone is insufficient to determine whether these fluctuations are driven by quantum criticality. This scenario is strongly motivated by the recent observation of quantum critical nematic fluctuations in the underdoped region of the Ba(Fe$_{1-x}$Co$_x$)$_2$As$_2$ phase diagram\cite{Worasaran*2020}.  Close to a quantum critical point the susceptibility ($\chi$) is anticipated to follow power law behavior both as a function of doping ($\lim_{T\to0} \chi\propto|x-x_c|^{-\gamma}$) and temperature ($\lim_{x\to x_c} \chi\propto T^{-\frac{\gamma}{z\nu}}$). Here $T$ is temperature, $x$ is doping, and $x_c$ is the doping at the quantum critical point. $\gamma$ and $z\nu$ are critical exponents that depend on the nature of the critical point.
	%To look for power law behavior the susceptibility needs to be measured at low temperatures close to the putative quantum critical point and for a fine spectrum of dopings. This is precisely the measurement we perform. 
Distance from the critical point, both in temperature and in doping, will introduce increasingly large corrections to the power law scaling.

\begin{figure}[t]
        \includegraphics[width=\columnwidth]{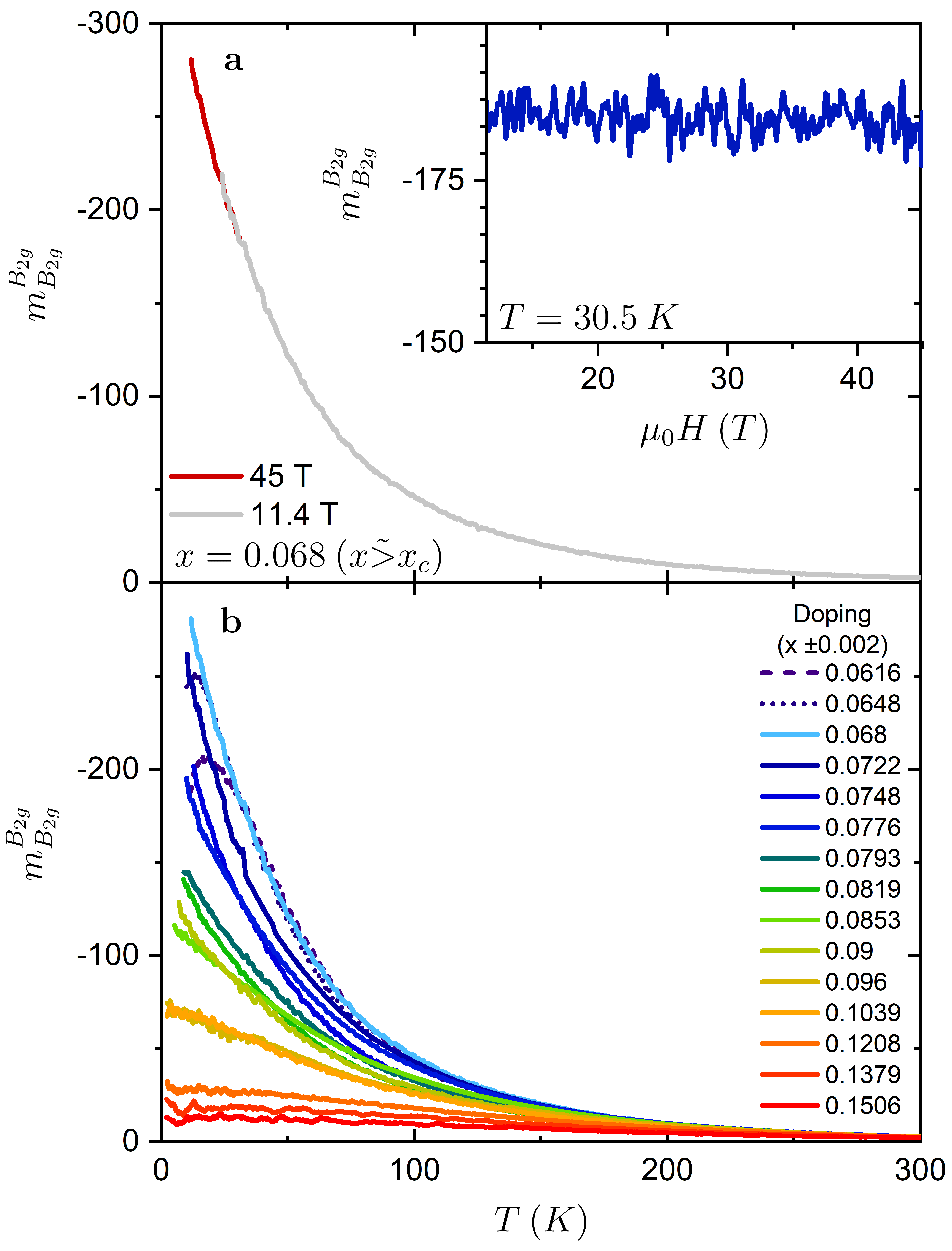}
        \caption{\textbf{The doping and temperature dependence of $m^{B_{2g}}_{B_{2g}}$.} \textbf{a}, Temperature dependence of $m_{B_{2g}}^{B_{2g}}$ for the composition closest to $x_c$, Ba(Fe$_{0.932}$Co$_{0.068}$)$_2$As$_2$, in fields of 11.4 T and 45 T. Inset shows the absence of any observable field dependence of $m^{B_{2g}}_{B_{2g}}$ above the zero field superconducting transition. \textbf{b}, The evolution of $m^{B_{2g}}_{B_{2g}}$ with doping as a function of temperature. Data taken with fields between 0 T - 45 T.
    	}
        \label{fig:sweeps}
\end{figure}

By symmetry the nematic susceptibility ($\chi_{B_{2g}}$) is related to a specific component of the elastoresistivity tensor\cite{Chu2012}, $m^{B_{2g}}_{B_{2g}}$ (the linear resistivity response ($\Delta\rho$) to shear strain ($\epsilon_{B_{2g}}$)) by a constant of proportionality ($g_{T,x}$),
\begin{equation}
\label{eqn:susc}
    \chi_{B_{2g}}=g_{T,x} \frac{(\frac{\Delta\rho}{\rho_0})_{B_{2g}}}{\epsilon_{B_{2g}}}= g_{T,x}m^{B_{2g}}_{B_{2g}}
\end{equation}
Here $\rho_0$ is the in-plane resistivity of the unstrained, tetragonal material. In practice we approximate $\rho_0$ with the $\epsilon_{B_{2g}}=0$ value of the resistivity, $\rho(\epsilon_{B_{2g}}=0)$. A more detailed and general description of this technique can be found in prior publications\cite{Chu2012, Shapiro2015, Shapiro2016, Kuo2016}. Previous measurements of $m^{B_{2g}}_{B_{2g}}$ for underdoped compositions reveal a Curie-Weiss functional form. Since this is the anticipated behavior for $\chi_{B_{2g}}$ approaching a thermally driven nematic phase transition, it was deduced that $g_{T,x}$ did not have an observable temperature dependence for those compositions.
%These previous works assume a temperature and doping independent constant of proportionality ($g_{T,x}$) between the nematic susceptibility and $m^{B_{2g}}_{B_{2g}}$ since they find a Curie-Weiss temperature dependence of $m^{B_{2g}}_{B_{2g}}$ for underdoped Ba(Fe$_{1-x}$Co$_x$)$_2$As$_2$. This is the expected functional form for the mean-field nematic susceptibility approaching a thermally driven nematic transition \cite{Chu2012}. We make the same approximation of a doping and temperature independent constant of proportionality throughout this manuscript. 
The fluctuation dissipation theorem relates the magnitude of the susceptibility, and thus by proxy the magnitude of $m^{B_{2g}}_{B_{2g}}$, to the strength of the equilibrium fluctuations. As a consequence, elastoresistivity is a very sensitive technique to probe the equilibrium electronic nematic fluctuations of the disordered state. Measurements close to the putative nematic quantum critical point, however, are complicated by the presence of superconductivity. Not only does superconductivity preclude resistance measurements but it also competes with and induces a back-bending of the structural transition\cite{Nandi2010}. Suppressing superconductivity in large magnetic fields removes %the back-bending of the structural transition 
the competition between superconductivity and the structural transition and permits resistivity measurements to considerably lower temperatures and for compositions much closer to the putative quantum critical point. The elastoresistivity response of Ba(Fe$_{1-x}$Co$_x$)$_2$As$_2$ has a negligible field dependence up to 45 T (Figure \ref{fig:sweeps}a), meaning that large magnetic fields are a small perturbation on the nematic fluctuations. %\textcolor{red}{\sout{This means that elastoresistivity measurements of $m^{B_{2g}}_{B_{2g}}$ in fields large enough to suppress superconductivity probe the same physics as zero field measurements of the normal state above the superconducting transition.}}

\begin{figure}
        \centering
        \includegraphics[width=\columnwidth]{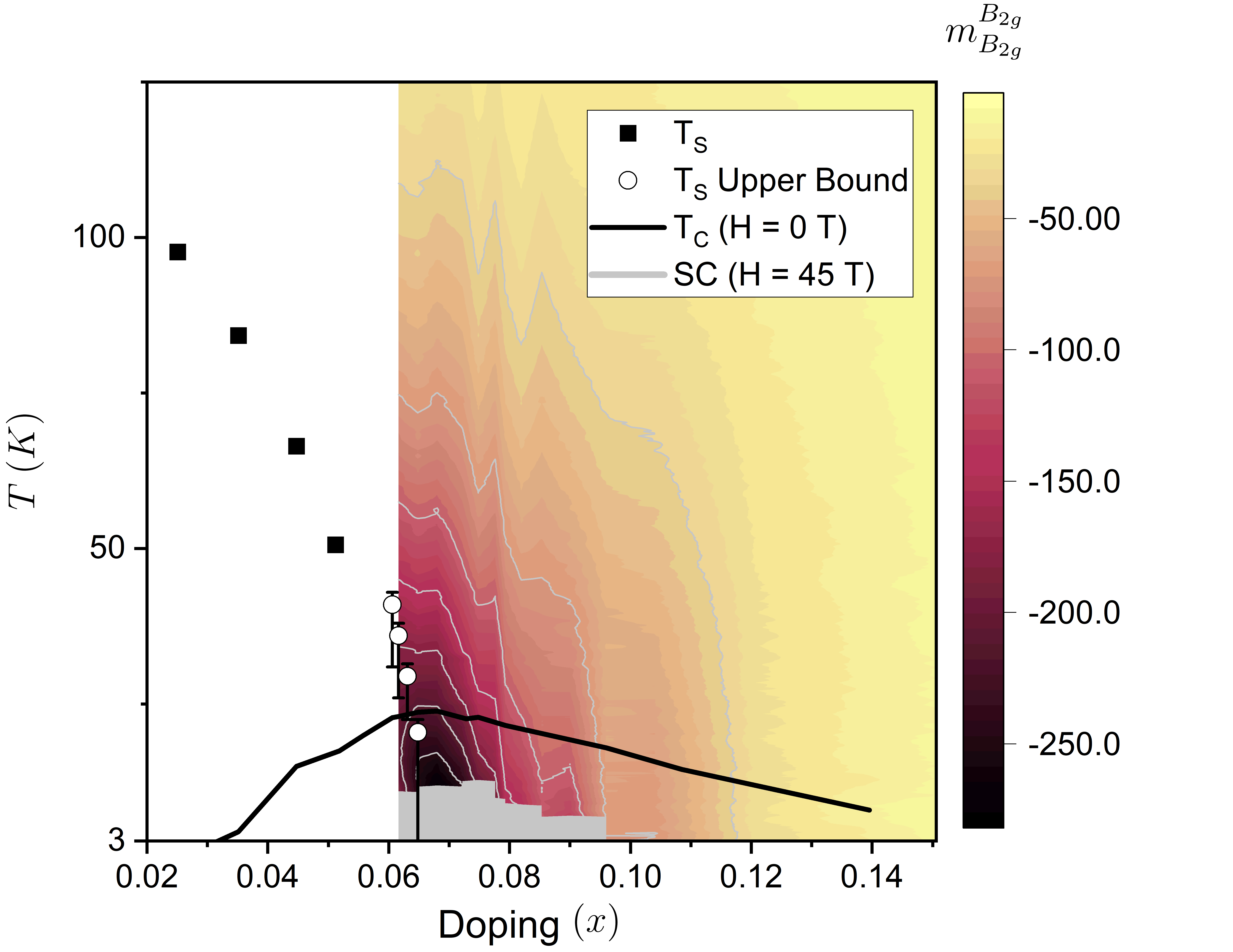}
        \caption{ \textbf{Phase diagram of Ba(Fe$_{1-x}$Co$_x$)$_2$As$_2$ overlaid with the doping dependence of $m^{B_{2g}}_{B_{2g}}$ for compositions with $x\geq0.0616$ (color plot).} The black line is the zero field superconducting transition and the gray region is the 45 T superconducting dome. The far underdoped structural transition temperatures (black squares) and zero field superconducting transition temperatures are from J.-H. Chu \textit{et al}.\protect\cite{Chu2009} The white circles represent the onset of the structural transition taken from resistivity measurements at 45 T (see Supplementary Information).}
        \label{fig:colorplot}
\end{figure}

\begin{figure}[h!t]
        \centering
        \includegraphics[width=\columnwidth]{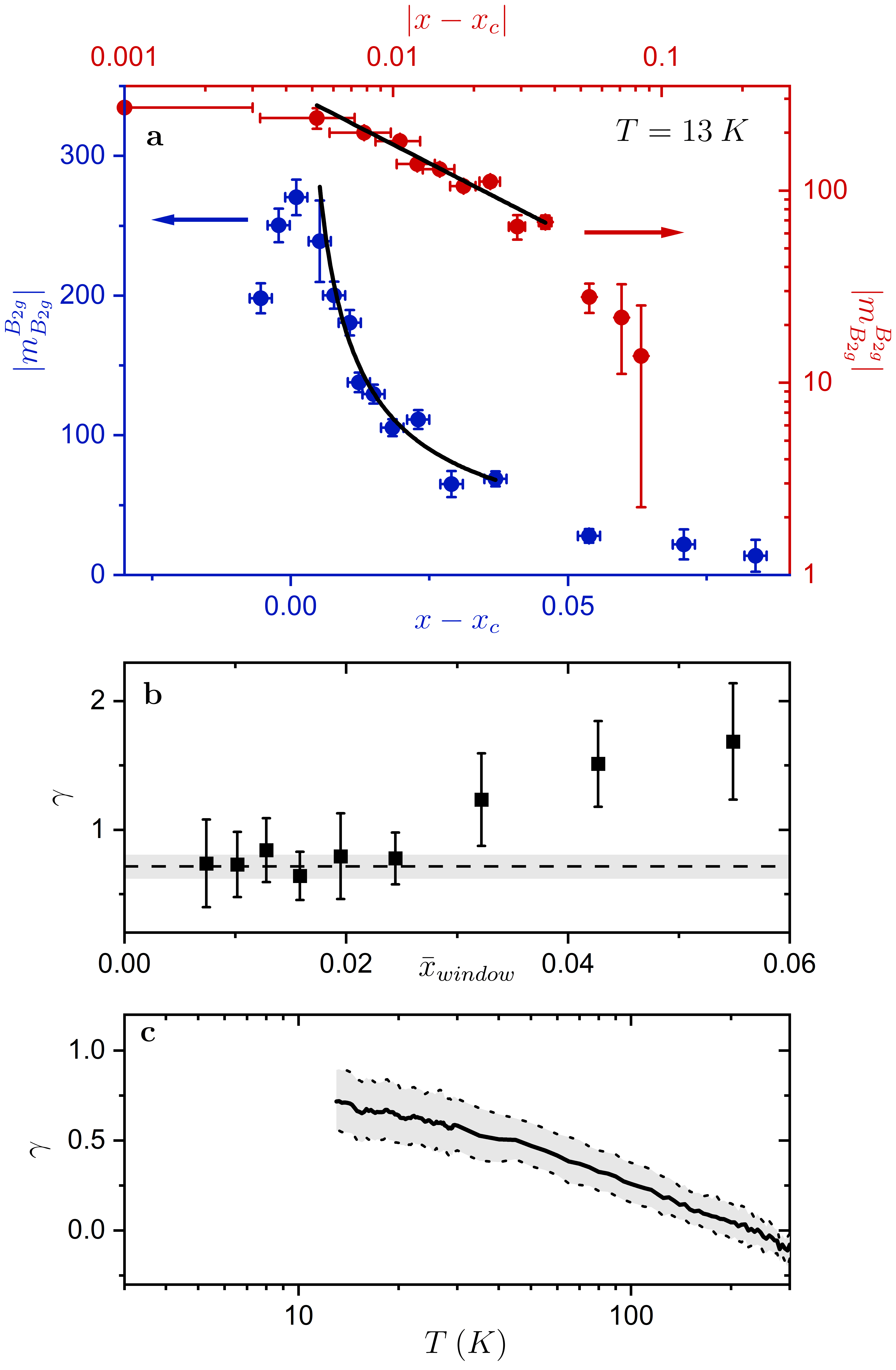}
        \caption{\textbf{Apparent power law behavior of $|m^{B_{2g}}_{B_{2g}}|$ as a function of $|x-x_c|$.} \textbf{a}, A linear (blue axes) and logarithmic (red axes) plot of $|m^{B_{2g}}_{B_{2g}}|$  vs $x-x_c$ at 13 K with power law fit $|m_{B_{2g}}^{B_{2g}}|\propto|x-x_c|^{-\gamma}$ (black lines). Error bars include the standard deviation of the measurement in addition to systematic errors. Additional details on included error available in the Supplementary Information. The fit was performed by fitting a line on the logarithmic plot for $0.0722 \leq x \leq 0.1039$ using the York computational method \cite{York2004}. \textbf{b}, The fitted critical exponent $\gamma$ for fits performed on a sliding 5 point window shown as a function of the average value of $x$ for the window. Overlaid on the plot is the extracted $\gamma$ from the fit performed in panel (a) (dashed line) and associated standard error (gray region), $\gamma= 0.72 \pm 0.09$. Error bars on each data point represent one standard error. Fits that do not include the three most overdoped samples all agree to within the standard error. \textbf{c}, The measured $\gamma$ (black line) as a function of temperature. Error (gray region) includes the standard error of the fits and error associated with uncertainty in the critical doping $x_c$ (see Supplementary Information).
    	}
        \label{fig:DopingCut}
\end{figure}

\begin{figure}[h!t]
        \centering
        \includegraphics[width=\columnwidth]{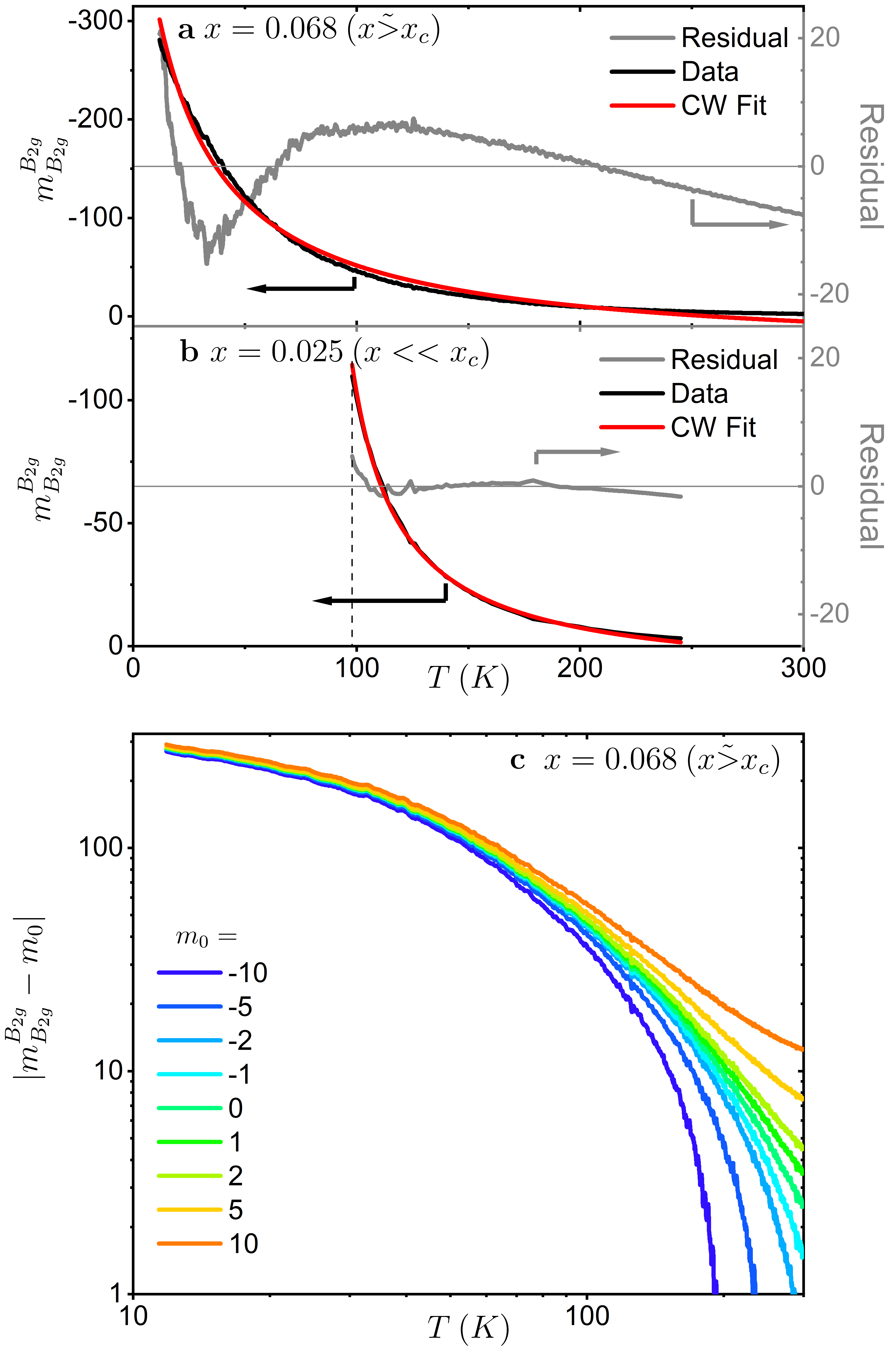}
        \caption{\textbf{The temperature dependence of $m^{B_{2g}}_{B_{2g}}$ for $x=0.068$ ($x\tilde{>}x_c$) cannot be described by a simple power law.} \textbf{a}, $m^{B_{2g}}_{B_{2g}}$ for $x= 0.068$ (black line), the best Curie-Weiss fit $m^{B_{2g}}_{B_{2g}}\propto\frac{\lambda}{a(T-\Theta)}+m_{0}$ (red line) and the associated residual (gray line). There is a clear temperature dependence in the residual indicating that Curie-Weiss does not fully describe the temperature evolution. \textbf{b}, $m^{B_{2g}}_{B_{2g}}$ for a far underdoped sample $x=0.025$ (black line), the best Curie-Weiss fit (red line) and the associated residual (gray line). The data and fit are taken from H.-H. Kuo \textit{et al}.\protect\cite{Kuo2016} This sample has a structural transition at 98 K (dashed line). The magnitude of the residual is small compared to the residual shown in panel (a) without a clear temperature dependence indicating that Curie-Weiss is a reasonable approximation of the functional form. \textbf{c}, Logarithmic plot of $|m^{B_{2g}}_{B_{2g}}-m_0|$ vs temperature for $x=0.068$ ($x\tilde{>}x_c$). No physically motivated value for $m_0$ linearizes the data.    	}
\label{fig:NoPLT}
\end{figure}

 Elastoresistivity measurements performed in large magnetic fields reveal that the magnitude of $m^{B_{2g}}_{B_{2g}}$ continues to smoothly increase with decreasing temperature in the absence of superconductivity and shows no evidence of saturation for compositions with $x\geq 0.068$. Underdoped samples ($x \lessapprox 0.067$)  exhibit a tetragonal-to-orthorhombic structural transition which coincides with or precedes a downturn in the elastoresistivity response (Figure \ref{fig:sweeps}b). The tetragonal-to-orthorhombic structural transition is suppressed with doping towards zero temperature near where $m^{B_{2g}}_{B_{2g}}$ is largest (Figure \ref{fig:colorplot}). The critical doping ($x_c$), \textit{i.e.} where the transition occurs at zero temperature, is estimated to be $0.067 \pm 0.002$ (see Supplementary Information) in the case where the structural phase transition remains second order. This composition marks the putative nematic quantum critical point in the absence of superconductivity.

The doping dependence of $m^{B_{2g}}_{B_{2g}}$ at 13 K (the lowest temperature where superconductivity can be suppressed for all dopings in 45 T) is shown in Figure \ref{fig:DopingCut}. There is an apparent divergence of $m^{B_{2g}}_{B_{2g}}$ upon approach to $x_c$ from the far overdoped side, with a maximum at $x= 0.068$ ($x\tilde{>}x_c$). Samples for $x \lessapprox 0.067$ are in the ordered phase at this temperature. To look for a power law dependence we plot the data for samples with $x\geq0.068$ on a logarithmic scale with log$|m_{B_{2g}}^{B_{2g}}|\propto -\gamma$log$|x-x_c|$ anticipated for scaling close to a quantum critical point. The reduced doping axis, $|x-x_c|$, spans nearly two decades from 0.001 to 0.0838 with 13 different compositions. Due to finite uncertainty in the measured doping concentration ($\pm 0.002$) there are large errors associated with the reduced composition ($|x-x_c|$) for the sample closest to the putative critical point ($x=0.068$) and it is excluded from fits. We performed a linear fit of the data for $x>0.068$ over a sliding 5-point window (Figure \ref{fig:DopingCut}b) using the York computational method to account for $x$ and $y$ errors \cite{York2004}. For windows that do not include the three most overdoped samples ($x\leq 0.1039$) the extracted slopes agree to within the standard error. The deviations seen for large dopings are consistent with the notion of increased corrections to the scaling function far from the critical doping. 

The above analysis indicates that  $m^{B_{2g}}_{B_{2g}}$ % appears to obey a power law dependence 
is consistent with a power law scaling versus $|x-x_c|$ for ($0.0722 \leq x \leq 0.1039$) which corresponds to nearly a decade in reduced doping (0.0052-0.0369). We cannot rule out other diverging functional forms, such as a lognormal distribution, however, since the power law dependence is the only theoretically motivated form it is the focus of this analysis. The temperature dependence of the extracted critical exponent, $\gamma$, is shown in Figure \ref{fig:DopingCut}c. The fitted $\gamma$ smoothly increases with decreasing temperature down to the lowest measured temperature (13 K) which corresponds to a $\gamma=0.72^{+0.18}_{-0.16}$. If $\gamma$ continues to smoothly increase, in the limit of $T \to 0$ K, $\gamma$ must be greater than this value. For reference, $\gamma=1$ is predicted for mean-field \cite{Sachdev2011a} and Hertz Millis \cite{Millis1993} quantum critical points. There is a small, temperature independent elastoresistivity response, $m_0$, which is expected to be on the order of the geometric factor. For the range of physically motivated values for $m_0$ the conclusions drawn here are robust and the extracted $\gamma$ at 13 K agree to within error with the $m_0=0$ fits shown in Figure \ref{fig:DopingCut}.

For underdoped compositions the temperature dependence of $m^{B_{2g}}_{B_{2g}}$ has been found to follow a Curie-Weiss functional form, $m_{B_{2g}}^{B_{2g}}=\frac{\lambda}{a(T-\Theta)} + m_0$\cite{Chu2012, Shapiro2015, Shapiro2016, Kuo2016}. Where $\frac{\lambda}{a}$ is the Curie constant, $\Theta$ is the Weiss temperature, and $m_0$ is the temperature independent elastoresistivity response. The temperature evolution of $m^{B_{2g}}_{B_{2g}}$ for both the sample closest to the critical doping, $x=0.068$ ($x\tilde{>}x_c$), and a far underdoped sample $x=0.025$ ($x<<x_c$) is shown in Figure \ref{fig:NoPLT} along with the best Curie-Weiss fits and residuals. The Curie-Weiss fit for the $x=0.068$ sample was performed over the whole temperature range with the best fit parameters $\frac{\lambda}{a}=-10960\pm36$, $\Theta=-20.3\pm0.1$, and $m_0=39.3\pm0.2$. The data and fit for the $x=0.025$ sample were taken from H.-H. Kuo \textit{et al}.\cite{Kuo2016} The fit was performed over a temperature window of 100 K - 205 K with best fit parameters $\frac{\lambda}{a}=-2706\pm32$, $\Theta=77\pm0.8$, and $m_0=14.5\pm0.8$. The low temperature cutoff is fixed by the structural transition. The fit for the $x=0.068$ sample not only has an unphysical value for the temperature independent response $m_0$, but the residual clearly has a systematic temperature dependence above the background measurement noise indicating that the data are not faithfully described by this functional form. In comparison, the residual for the underdoped $x=0.025$ sample is small and any temperature dependence in the residual is masked by the measurement noise. Over smaller temperature windows the data for the $x=0.068$ sample can be well fit by Curie-Weiss, but the data for the low temperature values of $m_{B_{2g}}^{B_{2g}}$ always fall below the divergence expected from Curie-Weiss behavior. This subCurie-Weiss behavior has been previously observed\cite{Kuo2016, Straquadine2019}, however here the measurements are performed over a larger temperature range and on a dense doping series through $x\tilde{>} x_c$ where the susceptibility, if driven by quantum critical fluctuations, is expected to be a power law in the clean limit. Theoretically effects due to weak disorder are predicted to suppress the divergence of the nematic susceptibility upon approach to a nematic quantum critical point \cite{Kuo2016} which is qualitatively consistent with the observed behavior.

At $x=x_c$ the susceptibility is expected to diverge at zero temperature, \textit{i.e.} if it was well-described by a Curie-Weiss functional form at the critical doping we expect $\Theta=0$. To look for power law behavior with any exponent in temperature we plot the data on a logarithmic scale with a range of physically motivated values for $m_0$ (Figure \ref{fig:NoPLT}c). Power law behavior would result in a linear response on the logarithmic plot, however, no value for $m_0$ linearizes the data. This indicates that the temperature dependence of the data for the $x=0.068$ ($x\tilde{>}x_c$) sample not only is not described by Curie's law, but in fact cannot by described by any single power law over the entire temperature range measured here. Additional attempted power law fits, including finite $\Theta$ values can be found in the Supplementary Information.

It is challenging to explain the dichotomy between the dependence of the nematic susceptibility on composition, where it is well described by a power law, and temperature, where the susceptibility deviates from power law behavior. A temperature and/or doping dependence in the constant of proportionality $g_{T,x}$ between the thermodynamic susceptibility $\chi_{B_{2g}}$ and the elastoresistivity coefficient  $m_{B_{2g}}^{B_{2g}}$ is insufficient to explain all the observations. A simple temperature dependent scaling could account for non-power law behavior in temperature. However, since previous measurements of optimally doped BaFe$_{2}$(As$_{1-x}$P$_x$)$_{2}$ reveal Curie-Weiss behavior down to the superconducting transition\cite{Kuo2016} and measurements on underdoped Ba(Fe$_{1-x}$Co$_x$)$_2$As$_2$ are consistent with power law behavior\cite{Kuo2016}, $g_{T,x}$ must also have a doping dependence. This scenario would imply breakdown of the power law scaling as a function of doping which would be inconsistent with the observation of power law behavior as a function of $|x-x_c|$. It is also possible that the corrections to the scaling relation are large for temperatures above 13 K (the practical lowest temperature accessible for compositions near optimal doping in fields of 45 T) or that the system evolves from a high temperature quantum critical regime where the energy scale of disorder is irrelevant to a dirty quantum critical regime at low temperatures. This would suggest that the temperature dependence of $m_{B_{2g}}^{B_{2g}}$ converges on a power law behavior at lower temperatures. Under this assumption, our data constrain any such low temperature power law with a best-fit estimate for the exponent of the temperature dependence of $\geq -0.33 \pm 0.12$ (see Supplementary Information). %\sout{Such a value is inconsistent with current theories for clean systems, although it is unclear what role disorder may play in theoretically expected critical scaling values.} 
Finally, it is possible that quantum criticality is not driving the strong nematic fluctuations in this material for these overdoped compositions. This leaves open the intriguing question of what is driving the apparent scale invariance in the doping dependence as witnessed by the power law variation of $\chi_{B_{2g}}$ as a function of $|x-x_c|$.

\section{Acknowledgments}
%The authors would like to thank Dr. Dale Burns for assistance with the 
The authors would like to thank A. Hristov, M. Ikeda, Y. Schattner, S. A. Kivelson, S. Raghu, and Q. Si for helpful discussions. We thank D. Graf and  A. Suslov for their assistance with measurements at the National High Magnetic Field Laboratory. EMPA measurements were done at the Stanford Microchemical Analysis Facility. A portion of this work was performed at the National High Magnetic Field Laboratory, which is supported by the National Science Foundation Cooperative Agreement No. DMR-1644779 and the State of Florida. This work was supported by the Department of Energy, Office of Basic Energy Sciences, under Contract No. DE-AC02-76SF00515. J.C.P. was supported by a Gabilan Stanford Graduate Fellowship and a Stanford Lieberman Fellowship.

\section{Author contributions}
J.C.P., P. W., and I.R.F. conceived of the experiment. J.C.P. and P.W. prepared and synthesized the samples. J.C.P. and D.B. performed the EMPA measurements. J. C. P., P. W., J. A. W. S., and M.E.S. performed the measurements in high magnetic fields with assistance from S.T.H. J.C.P. performed the data analysis with guidance from P.W. and I.R.F. J.C.P. and I.R.F. wrote the paper. All authors contributed to editing the manuscript.

\section{Methods}
Bulk single crystal samples were grown by using a self-flux technique described in detail in J.-H. Chu \textit{et al}. \cite{Chu2009} The Co-doping was measured for all material batches using electron microprobe analysis (EMPA). The parent compound BaFe$_2$As$_2$ and cobalt metal were used for calibration. Doping variation within a sample and within a batch were found to be characterized by a standard deviation of less than $0.002$.

Bulk samples were cleaved into square plates with in-plane dimensions $\geq$750 $\mu m$ and out-of-plane dimensions $\leq$40 $\mu m$. The samples were cut such that the edges were parallel to the tetragonal [110] direction. Gold pads were deposited on the corners of the samples using plasma sputtering and an aluminum foil mask. Electrical connection was made by dipping gold wires into EPO-TEK H20E conductive silver epoxy and adhering them onto the gold pads. The resistance of this setup is dominated by the gold wires and typical resistances are $\leq$ 3 $\Omega$. This modified Montgomery configuration allows for resistivity measurements simultaneously along the tetragonal [110] and [1$\bar{1}$0] directions.

Stress was applied to the samples by gluing them onto piezoelectric stacks (Part No.: PSt150/5x5/7 cryo 1, from Piezomechanik GmbH) with Masterbond EP21TCHT-1 epoxy. The samples were glued such that the edges were parallel to the edges of the piezoelectric stack (PZT) and the sample was submerged in epoxy with only a thin layer between the sample and PZT. Two samples were glued onto the front PZT face and a bi-directional resistive strain gauge (Micro-Measurements WK-06-062TT-350) was glued onto the back face of each PZT. The PZT was then mounted such that the applied magnetic field was perpendicular to the $ab$-plane of the samples. Two PZT stacks, with compositions $x = 0.0722$, 0.0853, 0.096, and 0.1208, detached from the probe wall during the experiment. The close 45 T superconducting transition temperatures to nearby compositions suggests that the possible misalignment of field from rotation of these stacks is minimal.

The PZTs were driven from a sine wave generated by a SR860 lock-in amplifier passed through a Tegam 2350 high voltage amplifier. The drive frequency was 23 Hz with an amplitude of 75 V$_{peak}$ for low temperature measurements and 50 V$_{peak}$ for high temperature measurements (typically the cooldown or low field temperature sweeps up to room temperature). Typical temperature sweep rates were 0.7 K/min for low temperature/high field measurements and 3 K/min for temperature sweeps up to 300 K. Current was sourced into the samples and strain gauges by a voltage controlled current source (CS580) which was driven from a sine wave generated by a SR860 lock-in amplifier. The current amplitude was 5 mA$_{RMS}$ and 1 mA$_{RMS}$ through the samples and strain gauges respectively. Typical current frequencies were 30-40 Hz for the samples and 200-400 Hz for the strain gauges. A heating test was performed at 8 K and heating was found to be $\leq$ 0.15 K for the maximum PZT drive and sample currents.

AC elastoresistivity measurements\cite{Hristov2018} were performed by directly locking into the side band using the dual mode of the SR860 lock-in amplifiers. A second SR860 for each channel was used to directly measure the average voltage. The strain gauges were measured through a Wheatstone bridge while the sample voltages were measured directly. A Savitzky-Golay filter with a 1 K window was used to remove background noise. Typically strain was measured along two orientations, parallel and orthogonal to the PZT poling axis. In some instances it was not possible to measure both strain gauges so the average measured Poisson ratio from all runs was used to calculate the overall strain. 

All samples were measured in the 45 Tesla Hybrid Magnet in a Helium-4 variable-temperature insert at the National High Magnetic Field Lab except for the two most overdoped samples, $x = 0.1379$ and $x = 0.1506$, which were measured in a 14 T PPMS made by Quantum Design. The measurements on the two overdoped samples were performed with a PZT drive voltage of 50 V$_{peak}$, a temperature sweep rate of 1 K/min, and filtered over a 4 K window.

\bibliographystyle{naturemag}
\bibliography{ReferencesHybridPaper}

\onecolumngrid
\pagebreak

\renewcommand{\figurename}{Supplementary Figure}
\setcounter{figure}{0}

\section{Supplementary Information}

\subsection{Experimental Errors}

Strain was measured with a bi-directional resistive strain gauge from Micro-Measurements (Part. No. WK-06-062TT-350). The strain gauges were measured in a Wheatstone bridge configuration with the three balance resistors at room temperature. Since strain was simultaneously measured along two orthogonal directions we were able to correct for the transverse strain sensitivity of each strain gauge using the manufacturer's provided calibration. We also accounted for the temperature dependence of the gauge factor. The largest uncertainties in the measurement of the oscillating strain experienced by the strain gauge are from balancing the bridge with the assumption that the line resistance of the cryostat was the thermalized value at either 30 K or 300 K depending on the temperature the bridge was balanced. Other smaller sources of error ($<0.6 \%$) include the thermal output of the strain gauge, the magnetoresistance of the gauge, and uncertainty in the manufacturer’s provided specifications. Overall error is estimated to be between 3$\%$-5$\%$, with a value of 4.1$\%$ used for the calculations in this manuscript. The measurement noise was quantified by taking a rolling standard deviation over a 1 K window (4 K for the two most overdoped samples). Additionally two samples, $x=0.0722$ and $x=0.096$, exhibited sharp shifts in the apparent measured value of $m^{B_{2g}}_{B_{2g}}$ on the order of ~10$\%$ which we tentatively attribute to changes in the current path through the silver paint contacts due to mechanical shifts during cooling. These three sources of error are included in the $y$-error bars in Figure \ref{fig:DopingCut}.
%The strain gauge was in series with the line resistance of the cryostat. For the meander most sensitive to strain along the poling axis of the PZT stack, SG$_{yy}$, the line resistance at 300 K was approximately 440 $\Omega$.

\subsection{Extracting x$_{c}$}
 \begin{figure}[h]
         \centering
         \includegraphics[width=0.5\columnwidth]{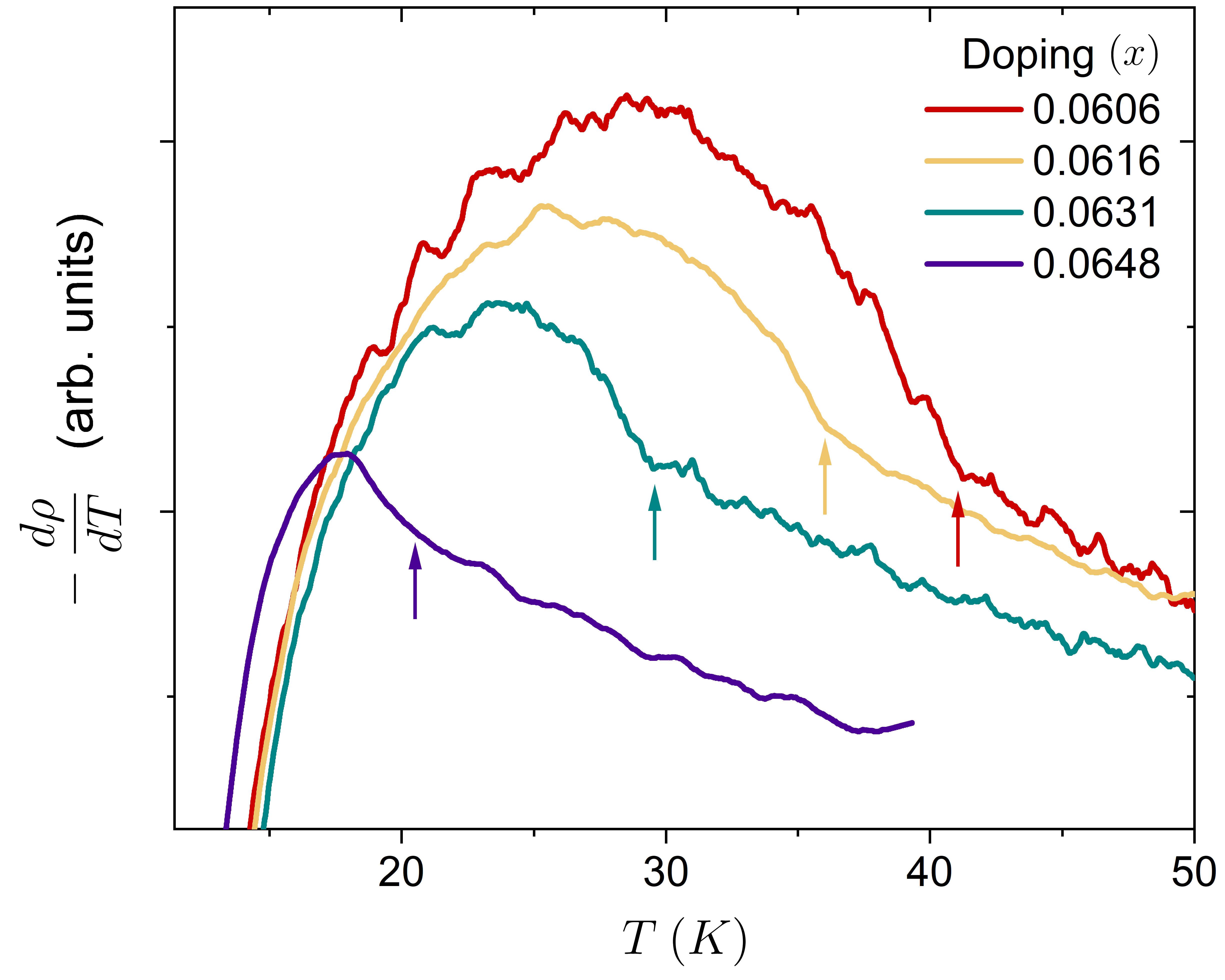}
         \caption{\textbf{The temperature derivative of resistivity for free standing samples with $x = 0.0606$, 0.0616, 0.0631, and 0.0648 in 45 T}. Vertical arrows mark the onset of a broad step in $-\frac{d\rho}{dT}$ and an upper bound in the structural transition temperature.
     	}
         \label{fig:dRdT}
 \end{figure}

In addition to the samples mounted for elastoresistivity measurements we performed four point resistivity measurements on free standing bar samples for $x = 0.0606$, 0.0616, 0.0631, and 0.0648. For far underdoped compositions ($x<0.051$) the structural transition can be identified by a mean-field like step in the temperature derivative of the resistivity ($\frac{d\rho}{dT}$). In zero field this feature broadens with doping and disappears above $x=0.051$\cite{Chu2009}. With the suppression of superconductivity in 45 T magnetic fields, signatures of the structural transition reappear in resistivity measurements. The features are still extremely broad and merge with the signature of the antiferromagnetic transition. From the temperature derivative of resistivity alone (Supplementary Figure \ref{fig:dRdT}) it is not possible to precisely determine the structural transition temperature. The onset of the structural transition can be bounded by the sharp change in the slope of $\frac{d\rho}{dT}$. Since there is no separation between the broad mean-field step associated with the structural transition and the onset of the antiferromagnetic order, the lower bound is set by the downturn of $-\frac{d\rho}{dT}$. This downturn for far underdoped compositions with sharp transitions occurs at temperatures below the subsequent antiferromagnetic transition. This is separate from the sharp downturn induced by the onset of superconductivity. For $x=0.0648$ there is no resolvable down turn associated with the antiferromagnetic transition before the onset of superconductivity.

\begin{figure}
         \centering
         \includegraphics[width=0.4\columnwidth]{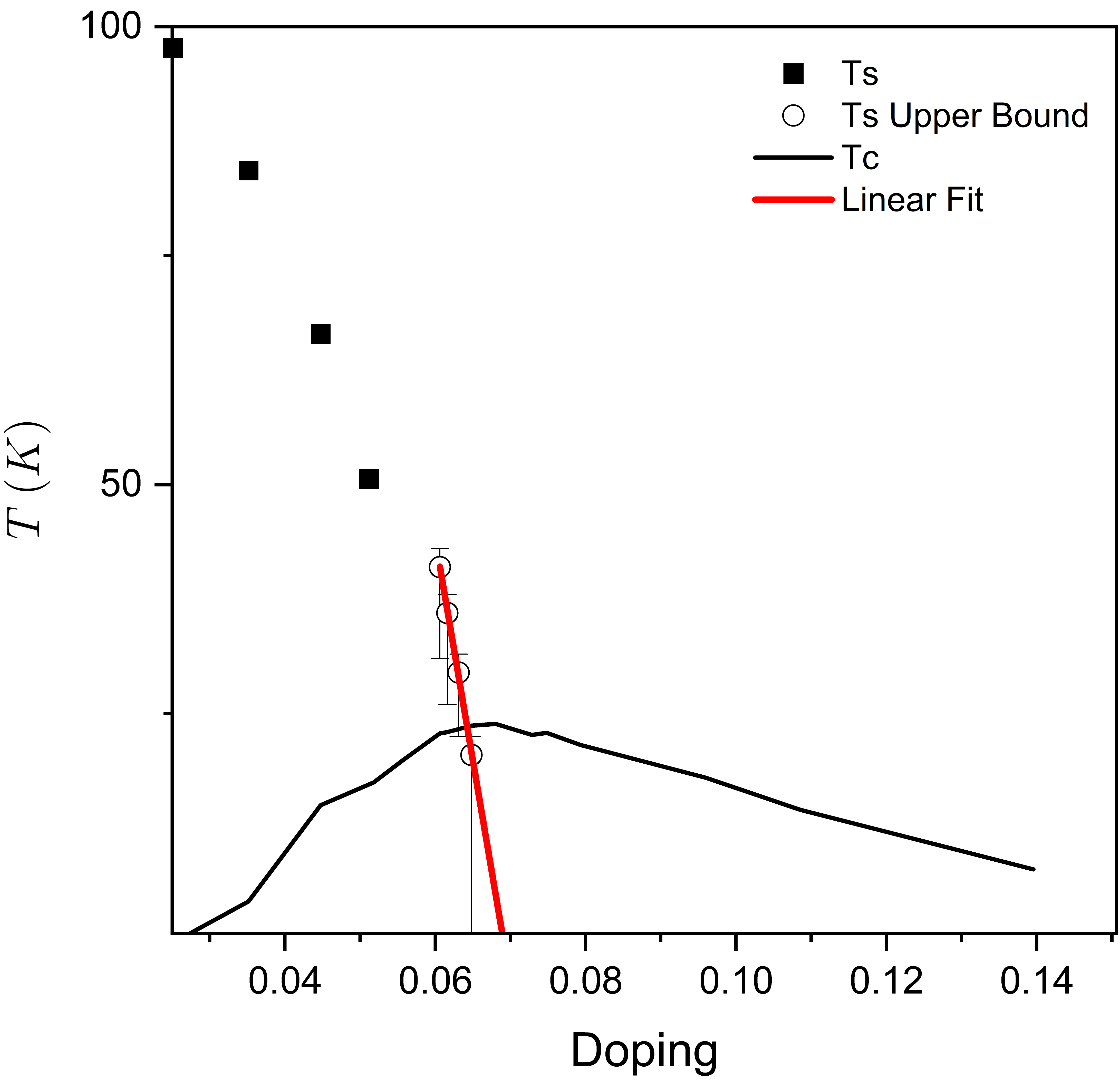}
         \caption{\textbf{The structural transition as a function of doping.}  The far underdoped structural transition temperatures (black squares) and zero field superconducting transition (black line) are from J.-H. Chu \textit{et al}.\protect\cite{Chu2009} while the white circles represent the onset of the structural transition taken from the resistivity measurements at 45 T shown in Supplementary Figure \ref{fig:dRdT}. A linear fit (red line) of $T_s(x)$ for the four samples closest to optimal doping ($0.0606 \leq x \leq 0.0648$) puts an upper bond on the critical doping of $x=0.069$. 
     	}
         \label{fig:xc}
 \end{figure}

The functional form of the structural transition versus doping is unknown, however we can set bounds on the critical doping $x_c$. The lower bound is set by the highest doping with observable signatures of the structural transition, $x = 0.065$. Since the phase transition is concave down, a conservative upper bound is a linear extrapolation of the structural transition vs doping (Supplementary Figure \ref{fig:xc}). A linear extrapolation of the structural transition for the four underdoped compositions closest to optimal gives an upper bound of $x = 0.069$. For this manuscript we use the value $x_c=0.067\pm0.002$.

\subsection{Fitting $\gamma$}

 \begin{figure}[h!]
         \centering
         \includegraphics[width=0.45\columnwidth]{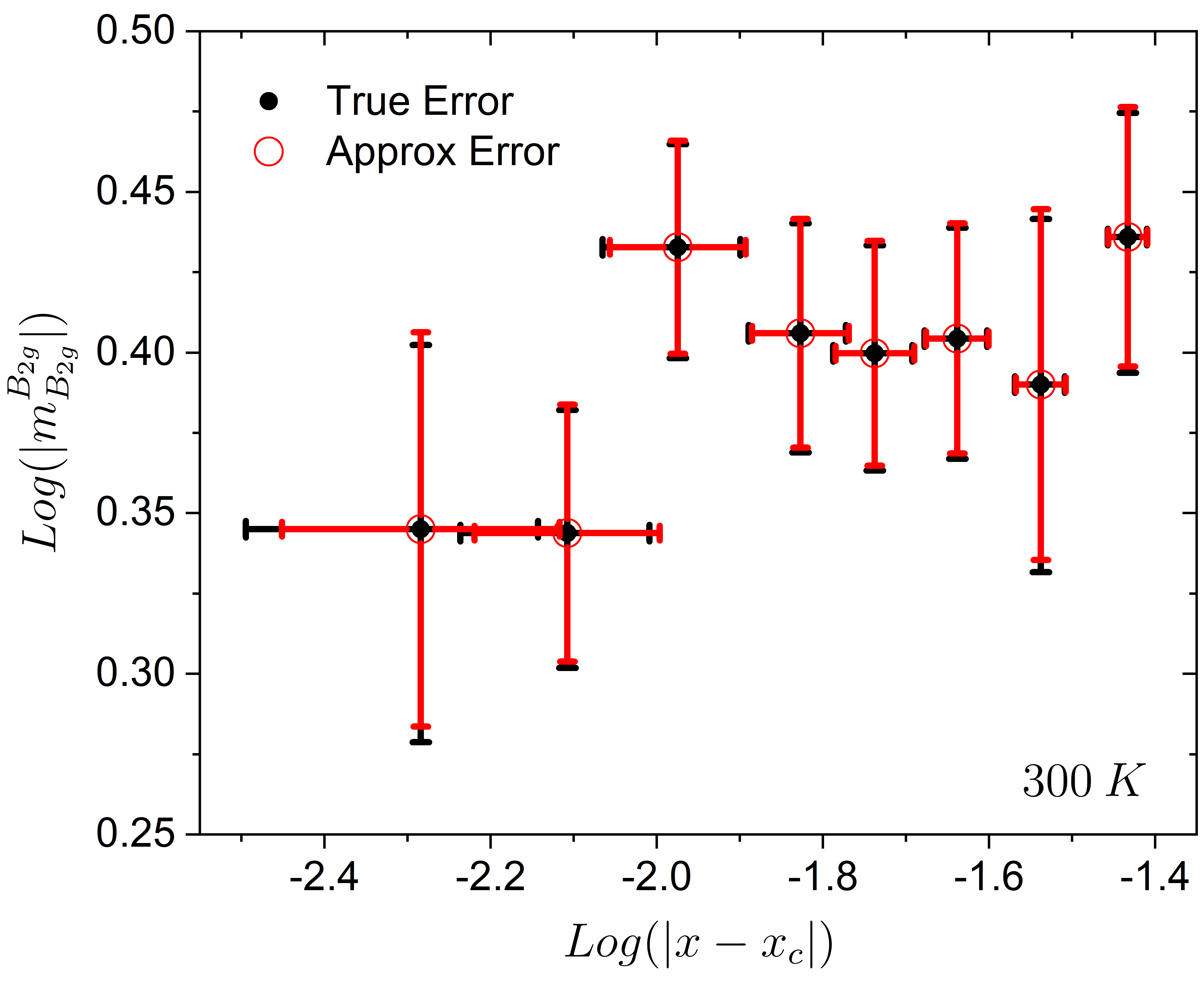}
         \caption{\textbf{Comparison of the actual measurement error of log($|m_{B_{2g}}^{B_{2g}}(x-x_c)|$) at 300 K versus a linear first order approximation.} The symmetric error approximation (red line) is a good approximation of the absolute error (black line) in the limit of small relative error on the linear scale.  
     	}
         \label{fig:errorapprox}
 \end{figure}

The critical exponent $\gamma$ is extracted from the slope of a linear fit of log($|m_{B_{2g}}^{B_{2g}}|$) versus log($|x-x_c|$). The uncertainty of the fit is an approximation assuming symmetric Gaussian errors (Supplementary Figure \ref{fig:errorapprox}) on the logarithmic scale using the York linear regression method \cite{York2004} with uncorrelated $x$ and $y$ errors (fits performed using the ``Linear Fit with X Error" analysis routine in Origin Pro 2019)\textsl{}. The error on the logarithmic scale ($\sigma_{log}$) can be computed from the symmetric error on the linear scale ($\sigma_x$) by $\sigma^{\pm}_{log}$=$|log(x\pm\sigma_x) - log(x)|$, however since log is not a linear function this produces asymmetric error bars. To calculate an approximation of the standard error of our fits in this paper we linearly approximate the measurement error as $\sigma_{log}\approx\frac{dlog(x)}{dx}\sigma_x=\frac{1}{ln(10)}\frac{\sigma_x}{x}$. A comparison of these errors are shown in Supplementary Figure \ref{fig:errorapprox}.

 \begin{figure}
         \centering
         \includegraphics[width=0.5\columnwidth]{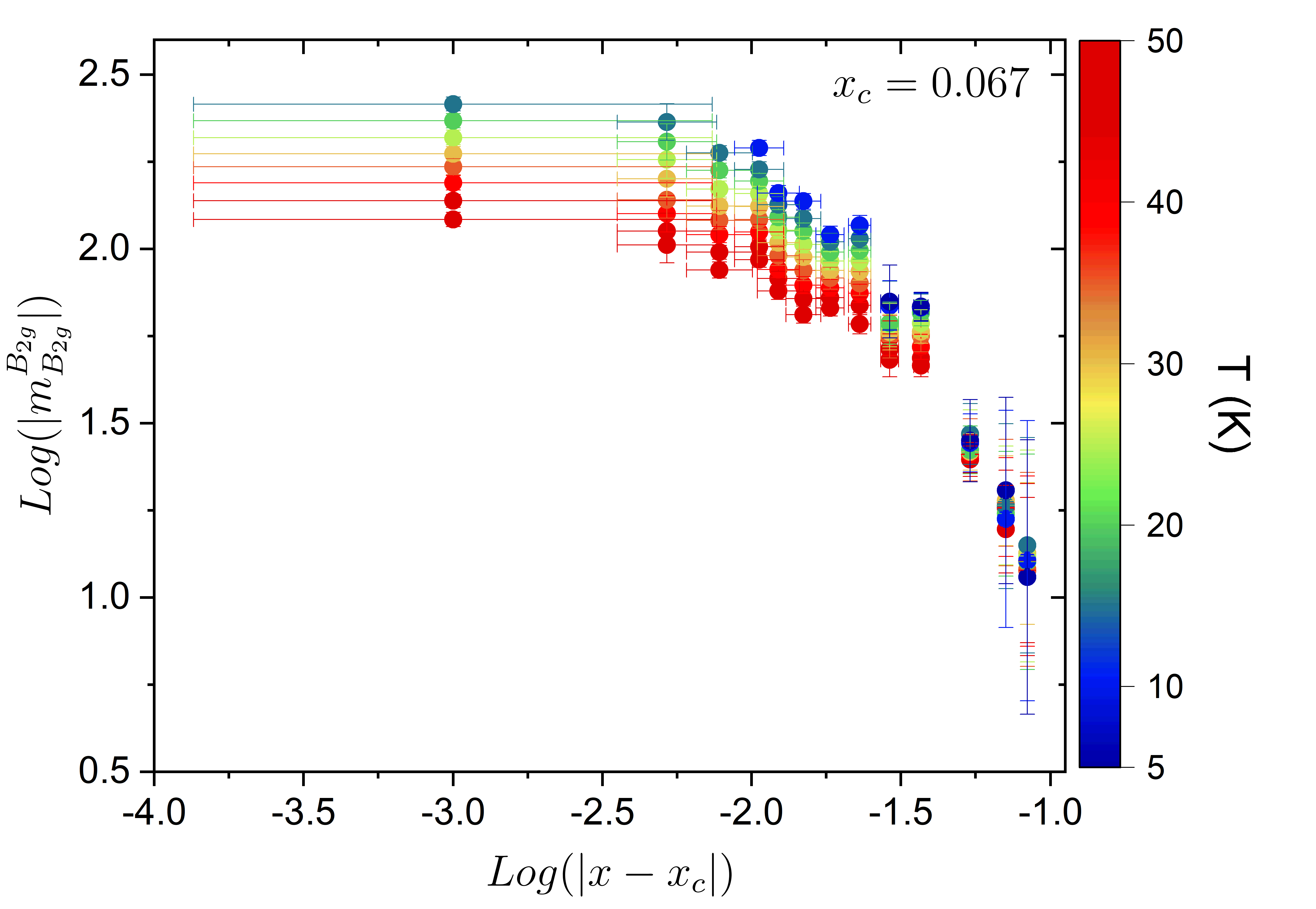}
         \caption{\textbf{Log($|m_{B_{2g}}^{B_{2g}}(x-x_c)|$) as a function of temperature.} Data are fixed temperature doping cuts from the temperature dependence shown in Figure \ref{fig:sweeps}b. Data are taken in fields between 0 T - 45 T with superconductivity fully suppressed. 
     	}
         \label{fig:mvsxTemp}
 \end{figure}

 \begin{figure}[h]
         \centering
         \includegraphics[width=0.5\columnwidth]{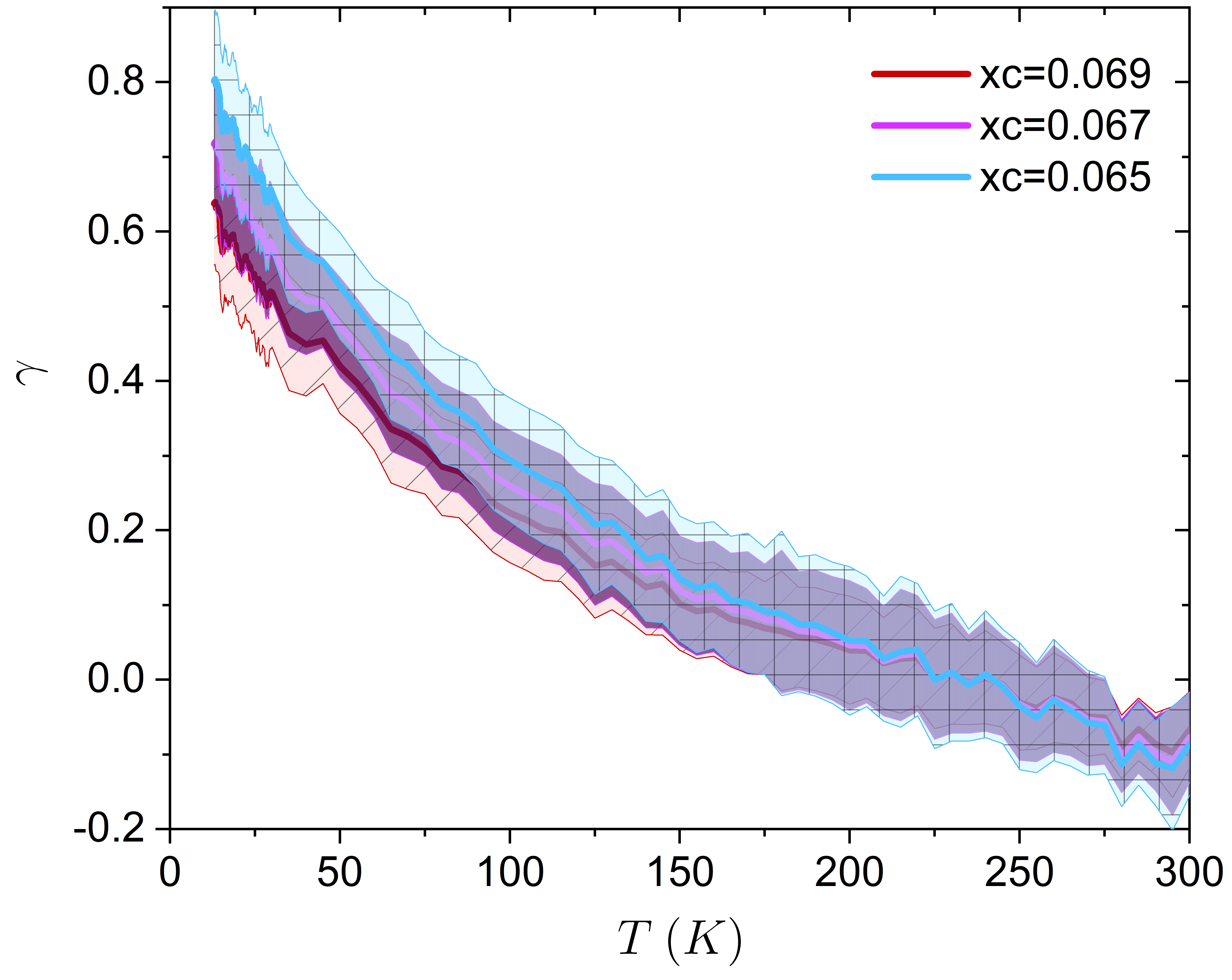}
         \caption{\textbf{The extracted $\gamma$ from a linear fit of log($|m_{B_{2g}}^{B_{2g}}(x-x_c)|$) vs log($|x-x_c|$) as a function of temperature for $0.068 <x \leq 0.1039$ and physically motivated values of $x_c$ with $0.065 \leq x_c \leq 0.069$.} The fitted $\gamma$ are displayed as solid lines and standard error for the fits are shown as shaded regions of the same color. Values for $\gamma$ presented in the text are for $x_c =0.067$. The quoted errors are the extremal error range from all fits with $0.065 \leq x_c \leq 0.069$.
     	}
         \label{fig:xcdep}
 \end{figure}

The fit was performed on dopings with $0.0722 \leq x \leq 0.1039$. This range was chosen since a linear fit over a rolling 5 point window is to within error constant within this range (Figure \ref{fig:DopingCut}b). For $x > 0.1039$ the fitted $\gamma$ begins to deviate. This is not unexpected as farther from the critical doping we expect large corrections to the scaling function. $x=0.068$ is not included in the fit since there is large $x$-error due to doping uncertainty in the reduced doping ($|x-x_c|$) axis. The magnitude of $m_{B_{2g}}^{B_{2g}}$ and $\gamma$ increase with decreasing temperature (Supplementary Figure \ref{fig:mvsxTemp}). 

Supplementary Figure \ref{fig:xcdep} shows the temperature dependence of $\gamma$ for $x_c=0.067\pm 0.002$. Below 13 K, 45 T is insufficient to fully suppress superconductivity for all dopings. The $\gamma$ value presented in the main text is taken using $x_c=0.067$ and the error bars are the extremal error range for values of $x_c$ between 0.065 to 0.069.

\pagebreak
	
\subsection{Curie-Weiss Fits of Ba(Fe$_{0.932}$Co$_{0.068}$)$_2$As$_2$}

 \begin{figure}[h!]
         \centering
         \includegraphics[width=0.5\columnwidth]{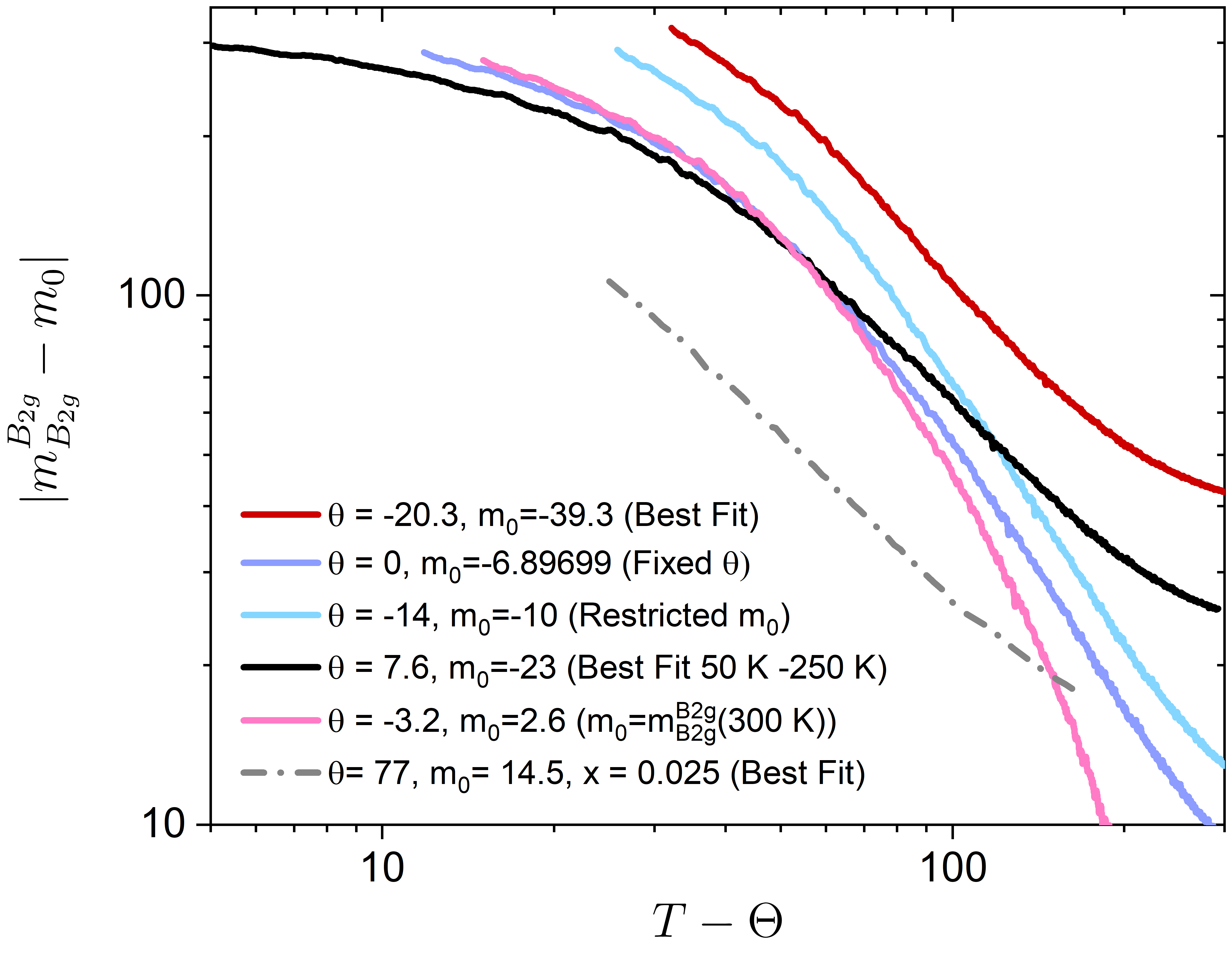}
         \caption{\textbf{Logarithmic plot of $|m_{B_{2g}}^{B_{2g}}-m_{0}|$ versus $T-\Theta$ for parameters motivated from Curie-Weiss fits of $m_{B_{2g}}^{B_{2g}}$ in Ba(Fe$_{0.932}$Co$_{0.068}$)$_2$As$_2$ (solid lines).} No physically motivated values of $m_0$ and $\Theta$ linearize the data over a wide temperature range. In comparison, the best Curie-Weiss fit of Ba(Fe$_{0.975}$Co$_{0.025}$)$_2$As$_2$ from H.-H. Kuo \textit{et al}. \protect\cite{Kuo2016} linearizes the data of the $x=0.025$ sample down to the structural transition at 98 K.     	}
         \label{fig:CW}
	\end{figure} 

Prior measurements of $m_{B_{2g}}^{B_{2g}}$ for far underdoped Ba(Fe$_{1-x}$Co$_x$)$_2$As$_2$ have shown that it can be well fit using a Curie-Weiss temperature dependence, $m_{B_{2g}}^{B_{2g}}$(T)=$\frac{\lambda}{a(T-\Theta)}+m_{0}$\cite{Chu2012, Shapiro2015, Shapiro2016, Kuo2016}. Here $m_0$ is the temperature independent background which is expected to be on the order of the geometric factor and $\Theta$ is the Weiss temperature. Near a quantum critical point $\Theta$ is expected to go through 0 K. 

For near optimally doped samples a subCurie-Weiss low temperature behavior has also been previously observed\cite{Kuo2016, Straquadine2019}. This is consistent with our measurements. A logarithmic plot of  $|m_{B_{2g}}^{B_{2g}}-m_{0}|$ versus $T-\Theta$ for parameters motivated by Curie-Weiss fits for Ba(Fe$_{0.932}$Co$_{0.068})_2$As$_2$ (i.e. $x\tilde{>}x_c$) are shown in Supplementary Figure \ref{fig:CW}. These include $m_0$ and $\Theta$ values for the overall best fit for the whole temperature range (red line), the best fit over the whole temperature range fixing the Weiss temperature to 0 K (purple line), restricting the temperature independent term to be on the order of the geometric factor (blue line), fitting Curie-Weiss over a restricted temperature range (black line) and fixing the temperature independent value to be the room temperature value (pink line). A linear slope of 1 on a logarithmic plot would be consistent with a Curie-Weiss behavior, however no fit linearizes the data over a wide temperature window and the low temperature data diverges at a slower rate than predicted from a Curie-Weiss form for all fits. This can be compared to the Curie-Weiss fit of Ba(Fe$_{0.975}$Co$_{0.025}$)$_2$As$_2$ from H.-H. Kuo \textit{et al}.\cite{Kuo2016} (gray dashed line) which is linear for a wide temperature window above the structural transition at 98 K. The Curie-Weiss fit over a restricted temperature range (black line) spans the same range in reduced doping (approximately three-quarters of a decade) as the fit performed on the underdoped sample.

\subsection{Low Temperature Power Law of m$_{B_{2g}}^{B_{2g}}(T)$}

$m_{B_{2g}}^{B_{2g}}(T)$ for $x\tilde{>}x_c$ does not follow a single power law as a function of temperature as shown by the nonlinear relationship of log($|m_{B_{2g}}^{B_{2g}}|$) vs log($T$) (Figure \ref{fig:NoPLT}c). If the temperature dependence is converging on power law behavior as the system is tuned towards the putative quantum critical point at zero temperature then $\frac{dlog(|m_{B_{2g}}^{B_{2g}}-m_0|)}{dlog(T)}$ should approach the power law value as the system is cooled. We approximate this derivative using the slope of the linear fit of log($|m_{B_{2g}}^{B_{2g}}-m_0|$) vs log($T$) over a rolling 1 K window for $m_0=-10$, 0, and 10. The result is shown in Supplementary Figure \ref{fig:lowTpower} along with the rolling mean over a 5 K window (gray line) and associated standard deviation of the mean (gray shaded region). The low temperature value at 15 K and standard deviation, -0.33$\pm$0.12, are presented in the main text. Overall the magnitude of the estimated power law exponent is decreasing with decreasing temperature. If this trend continues in the low temperature limit the exponent must be $\geq$ -0.33$\pm$0.12. 

 \begin{figure}[h!]
         \centering
         \includegraphics[width=0.5\columnwidth]{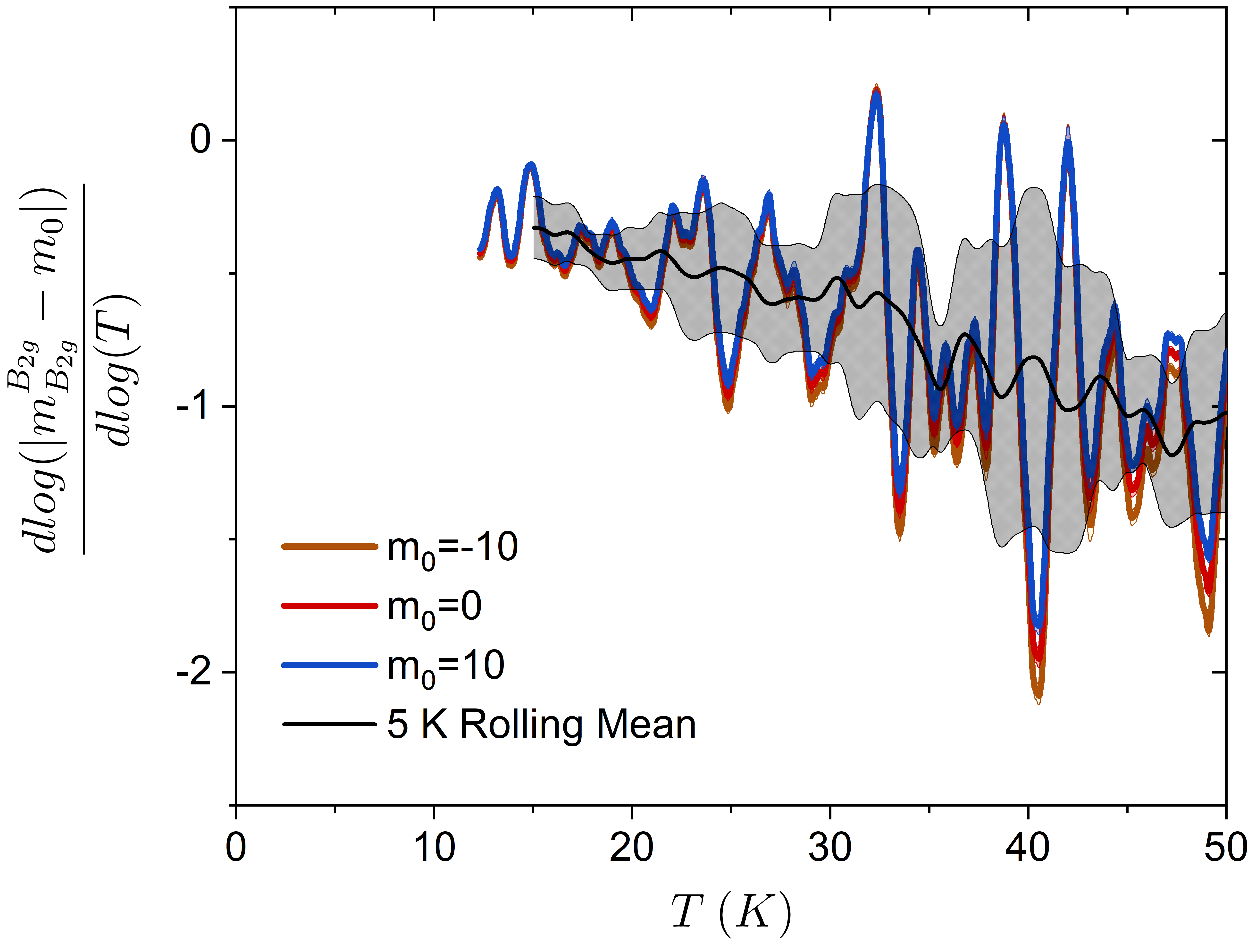}
         \caption{\textbf{Estimation of the power law exponent of the nematic susceptibility as a function of temperature assuming the system is converging on power law behavior in temperature as $T \to 0$ K. } Colored lines are $\frac{dlog(|m_{B_{2g}}^{B_{2g}}-m_0|)}{dlog(T)}$ calculated over a 1 K window for $x = 0.068$ and $m_0=-10$, 0, and 10. Gray line is the mean calculated over a rolling 5 K window. Error (gray shaded region), standard deviation of the mean.}
         \label{fig:lowTpower}
	\end{figure} 

\end{document}